\documentclass[preprint,aps,prb,groupedaddress,amssymb,showpacs]{revtex4}%
\usepackage{dcolumn}
\usepackage{amsmath}
\begin{document}

\title{Negative density of states: screening, Einstein relation, and negative diffusion. }
\author{A. L. Efros}
\email{efros@physics.utah.edu} \affiliation{Department of Physics,
University of Utah, Salt Lake City UT, 84112 USA}


\begin{abstract}
In strongly interacting electron systems with low density and at
low temperature the thermodynamic density of states is negative.
It creates difficulties with understanding of the Einstein
relation between conductivity and diffusion coefficient. Using the
expression for electrochemical potential that takes into account
the long range part of the Coulomb interaction it is shown that at
negative density of states Einstein relation gives a negative sign
of the diffusion coefficient $D$, but under this condition there
is no thermodynamic limitation on  the sign of $D$. It happens
because the unipolar relaxation
 of inhomogeneous electron density is not described by the diffusion equation.
 The relaxation  goes much faster due to electric forces caused by electron density
 and by neutralizing background. Diffusion coefficient is irrelevant in this case and it
 is not necessarily positive  because process of diffusion does not contribute
 to  the positive production of entropy.
 In the case of bipolar diffusion negative $D$ results in a global
  absolute instability that leads to formation of neutral excitons.
 Graphene is
     considered as an example of a system, where the density relaxation is expected
     to be due to electric force rather than diffusion. It may
     also have a negative density of states.
\end{abstract}
\pacs{71.27. +a,73.50.-h} \maketitle

\section{Introduction}

The idea of the Einstein relation was put forward  by
Einstein\cite{ALB} and Smoluchowski\cite{SM} in 1905-1906. Both
scientists considered the Brownian motion in the presence of
gravitational force.
 The result is the relation between mobility $u$ in the field and diffusion coefficient $D$.
 In  case of electric field and particles with the charge $e$ it has a form
\begin{equation}\label{u}
    eD=Tu,
\end{equation}
where $T$ is the temperature in energy units. The main idea was
equivalence of an  external force and the density gradient. Of
course, both Einstein and Smoluchowski did not care about
negligible mutual gravitational or any other small interactions of
the Brownian particles.

The  formulation of the Einstein relation for
electrons  is based upon electrochemical potential, the thermodynamic function that,
like temperature and pressure, should be the same at all points of the system in the
 equilibrium state. The usual arguments are as follows. If an external potential $\psi$
 is applied to the system, the condition of thermodynamic  equilibrium reads
\begin{equation}\label{ec}
  E_{ec}=\mu (n)+e\psi=Const,
\end{equation}
where $\mu(n)$ is the chemical potential as a function of inhomogeneous electron density $n$.
In the equilibrium both $n$ and $\psi$ are function of coordinates while $E_{ec}$ is constant.
 The temperature $T$ should also be constant. Therefore, the electrical current density ${\bf j}$ at
 constant $T$ can be written in a form\cite{LL}
\begin{equation}\label{j}
{\bf j}=-\frac{\sigma}{e}\nabla E_{ec}=\sigma {\bf E}-D \nabla en,
\end{equation}
where $\sigma$ is conductivity and ${\bf E}=-\nabla \psi$. Then one gets relation connecting $\sigma$ and $D$
\begin{equation}\label{Ein}
\frac{\sigma}{e^2} \frac{d\mu}{dn}=D,
\end{equation}
which is also called Einstein relation. For the Boltzman gas $d\mu/dn=T/n$ and one gets Eq.(\ref{u}) if $\sigma=enu$.
It looks like derivation of Eq. (\ref{Ein}) is independent of the properties of the system and this equation
 can be consider as general thermodynamic law.

A simple observation shows however that in the case of non-ideal
electron gas the Einstein relation needs some comments. We discuss
 an electron gas on the positive background at low temperatures and low densities when dimensionless parameter
 $r_s$ is not very small. Here $r_s^3=3/(4\pi n a_B^3)$ for 3-d case and
 $r_s^2=1/\pi n_2a_B^2$,  where $n$ and $n_2$ are 3- and 2-dimensional
 electron densities respectively and $a_B=\hbar^2 \kappa/m e^2$ is
 the  Bohr radius, $m$ is  an effective electronic mass,
 $\kappa$ is an effective permittivity.

  The problems of
  dynamic screening and diffusion in slightly non-ideal electron
  gas ($r_s<<1$) with electron-electron interaction were
considered in details about 20 years ago (See Ref.[\cite{alt, zu,
 alt1}])In this case the thermodynamic density of states is large
 and positive. I concentrate here on the strongly non-ideal case $r_s\geq 1$.

 An electron gas on the positive background at low temperatures and low
  densities has energy $E$ of the order of $-e^2 n^{1/d}N/\kappa$,
  where $d=2,3$ is the space dimensionality and $n$ is the density per
  area or volume respectively, $N$ is total number of electrons.
  Then $\mu\sim -e^2 n^{1/d}/\kappa$ and $E,\mu$, and $d\mu/dn$
  are negative\cite{BE,SE}. The first experimental confirmation of
  this idea was done by Kravchenko {\it et al}\cite{KRL, KR}, but
  direct quantitative study of this effect was performed by Eisenstein {\it et al}\cite{EI,
  EI1}.

 The derivative  $d\mu/dn$ is proportional to the
reciprocal compressibility of the electron gas. Note that compressibility has to
be positive due to
the thermodynamical condition of stability. However, this principle cannot be applied to the
charged systems, like electron gas, because part of their energy
is outside the system in a form of the energy of  electric field.
On the other hand, in the case of a neutral electron-hole plasma,
the situation of negative compressibility can arise leading to
collapse of the system. Such a situation is considered at the end
of Sec. III.

It follows from Eq. {\ref{Ein} that if $d\mu/dn$ is negative,
diffusion coefficient $D$ and conductivity $\sigma$ have opposite
signs. This observation needs an explanation  because near the
thermodynamic equilibrium both of them have to be positive to
provide positive entropy production due to the Joule heat and due
to the relaxation of inhomogeneous density.
\section{Electrochemical potential and static screening}

To resolve   this contradiction one should include the long-range
part of the Coulomb potential created by inhomogeneous electron
gas into the function $E_{ec}$ in Eq. (\ref{ec}). This
contribution is a functional of $n({\bf r})$.

To find $E_{ec}$ taking into account electron-electron interaction one should minimize
 the Helmholtz energy $F$ with respect to electron density $n({\bf r})$ at a given  value of $T$ and
 $N$. For low $T$ one gets
\begin{eqnarray}
\nonumber
F=\frac{e^2}{2\kappa}\int\int\frac{n^\prime({\bf r})n^\prime({\bf r^\prime})d^3 r d^3 r^\prime}{|{\bf r-r^\prime}|}+
\int f(n+n^\prime)d^3 r+
\nonumber
\\
\int e n^\prime({\bf r})\psi d^3 r-E_{ec}\int n^\prime({\bf r})d^3 r,
\label{F}
\end{eqnarray}
where $f$ is the Helmholtz energy density of a homogeneous
electron system that results from  the interaction in a neutral
system, like  the Wigner crystal or "Wigner liquid". Since this
interaction comes mainly from the nearest neighbors and $n({\bf
r})$ is a smooth function, one may assume that both $f$ and
chemical potential $\mu=df/dn$ are  local functions of $n({\bf
r})$
 We assume also that $n({\bf
r})=n+n^\prime({\bf r})$, where $n$ is average density and
$n^\prime\ll n$.

Minimization of this expression with respect to $ n^\prime$ gives the equation
\begin{equation}\label{ec2}
E_{ec}=\mu(n)+e\psi
+\frac{d\mu}{dn}n^\prime+\frac{e^2}{\kappa}\int\frac{n^\prime({\bf
r^\prime}) d^3 r^\prime}{|{\bf r-r^\prime}|}.
\end{equation}

It differs from Eq. (\ref{ec}) by the potential of electrons in
the right hand side. Note that this potential is due to the
violation of neutrality in a scale much larger than the average
distance between electrons.
 To check
this equation we consider thermodynamic equilibrium and find
equations for the Thomas-Fermi static screening in 3- and
2-dimensional cases. Since  $E_{ec}$ is independent of ${\bf r}$
in thermodynamic equilibrium one may take
  $E_{ec}-\mu(n)$ as
 a reference point for the total potential $\varphi$ defined as
\begin{equation}\label{pot}
    \varphi=\psi+\frac{e}{\kappa}\int\frac{n^\prime({\bf r^\prime}) d^3 r^\prime}{|{\bf
    r-r^\prime}|}.
\end{equation}
It follows from Eq. (\ref{ec2}) that
\begin{equation}\label{1}
e\varphi=-\frac{d\mu}{dn}n^\prime.
\end{equation}
The Poisson equation has a form
\begin{equation}\label{p3}
\nabla^2\varphi=-\frac{4\pi (e n^\prime-\rho_{ext})}{\kappa},
\end{equation}
where $\rho_{ext}$ is density of external charge.
Using Eq. (\ref {1}) one gets final equation for the 3-d linear screening
\begin{equation}\label{3s}
 \nabla^2\varphi=-q_3^2\varphi- \frac{4\pi \rho_{ext}}{\kappa}.
\end{equation}
Here
\begin{equation}\label{3r}
  q_3^2=\frac{4 \pi e^2}{\kappa}\frac{dn}{d\mu}
\end{equation}
is the reciprocal 3-dimensional screening radius.

Consider now a thin layer (x-y plane) with 2d electron gas
separating two media with dielectric constants $\kappa_1$ and
$\kappa_2$. In this case one should substitute $n\Rightarrow
n_2\delta(z)$ and
 $\kappa\Rightarrow \bar{\kappa}=(\kappa_1+\kappa_2)/2$. The results is\cite{ando}
\begin{equation}\label{2s}
 \nabla^2\varphi=-q_2\varphi \delta(z) - \frac{4\pi \rho_{ext}}{\bar{\kappa}},
\end{equation}
where
\begin{equation}\label{2r}
  q_2=\frac{2 \pi e^2}{\bar{\kappa}}\frac{dn_2}{d\mu}.
\end{equation}
It is important that Eqs. (\ref{3s}), (\ref{2s}) are applicable
only if the screening is linear ($n^\prime\ll n$)\cite{ef1}. There
is another serious problem of applicability the Thomas-Fermi
approximation in the case of the negative density of states.
Indeed, the dielectric permittivity in this approximation has a
form
\begin{equation}\label{ep3}
\epsilon(q)=\kappa(1-\frac{|q_3^2|}{q^2})
\end{equation}
in 3-d case and
\begin{equation}\label{ep2}
\epsilon(q)=\kappa(1-\frac{|q_2|}{q})
\end{equation}
in 2-d case. In both cases it has  roots at $q=|q_3|,|q_2|$.  The
expression for the screened potential $\varphi$ has a form
\begin{equation}\label{scr}
\varphi({\bf r})=\int \frac{\varphi_0({\bf q})\exp(i {\bf q \cdot
r})d{\bf q}}{\epsilon({\bf q})},
\end{equation}
where $\varphi_0$ is a bare potential. Thus, the roots of
$\epsilon$ transform into the first order poles without any
reasonable way of the detour. Such a detour follows from the
casuality for the $\omega$-plane but not for the q-plane.
Moreover, the electrostatic potential should be real and one
cannot add a small imaginary part in the denominator. Therefore I
think that the poles do not have any physical sense.

The reason is that negative sign of the density of states appears
when $q_3,q_2$ are of the order of average distance between
electrons $\bar{r}$. At such distances the very concept of
macroscopic field does not have sense. However, if the bare
potential has only harmonics with $q\ll |q_3|,|q_2|$, the
Eqs.(\ref{3s},\ref{2s}) have a sense. Consider, for example, the
screening of the positive charge $Z$ at a distance $z_0$ from the
plane with 2-d gas(plane $z=0$. The solution of Eq.(\ref{2s}) has
a form\cite{ando}
\begin{equation}\label{be}
\varphi(\rho)=\int_0^\infty \frac{Z \exp(-q z_0)}{\kappa(q+q_2)}q
J_0(q \rho)dq,
\end{equation}
where $\rho$ is a polar radius in the plane $z=0$. Suppose that
$|q_2|z_0\gg 1$. Now the contribution to integral Eq.(\ref{be})
from $q\simeq |q_2|$ is exponentially small and one can ignore $q$
in the denominator. Then
\begin{equation}\label{Z}
\varphi(\rho)=\frac{Z z_0}{\kappa
q_2\left(z_0^2+\rho^2\right)^{3/2}}.
\end{equation}
Note that at $q_2<0$ a positive charge creates a small {\em
negative} potential in the plane with electrons. That is what I
call "overscreening".

Extra electron density, as calculated from Eq. (\ref{1}) is
\begin{equation}\label{den}
e n^{\prime}=-\frac{Z z_0}{2\pi(z_0^2+\rho^2)^{3/2}}
\end{equation}
It is negative and independent of the sign of $q_2$. One can see
that the total charge
\begin{equation}\label{int}
\int_0^{\infty} e n^{\prime}2\pi\rho d \rho=-Z
\end{equation}
Due to geometry of the problem electric field  is zero below the plane with
electrons. As follows from Eq. (\ref{1}), the signs of
charge density and potential are opposite
 if the density of states is negative.

For the case of two such planes (double quantum well structure)
 Luryi\cite{SL} has  predicted  a small penetration of
electric field through the first plane. He has considered the case
 of  positive density of states. Then the small
penetrating field between two planes has the same direction as the
incident field.

Eisenstein {\it at al.}\cite{EI} studied this effect
experimentally and found out that at negative density of states
the propagating field is opposite to the incident field and this
is also a result of the overscreening (see the quantitative theory
in Ref.\cite{BU, PE, EI1}).

Negative density of states was also used\cite{AL} for the explanation of
magnetocapacitance data by Smith {\it at al.}\cite{S}.

\section{Conductivity versus diffusion}
 Now I come back to the problem of the negative diffusion. If
the system is not in equilibrium the electric current can be
written in the same form as Eq. (\ref{j})
\begin{equation}\label{j2}
{\bf j}=-\frac{\sigma}{e}\nabla E_{ec}.
\end{equation}
Using Eq. (\ref{ec2}) one gets
\begin{equation}\label{j2}
{\bf j}=\sigma {\bf E}-D\nabla e
n^\prime-\sigma\frac{e}{\kappa}\nabla\int\frac{n^\prime({\bf
r^\prime}) d^3 r^\prime}{|{\bf r-r^\prime}|}.
\end{equation}
Here D is connected to $\sigma$ by the Einstein relation Eq.
(\ref{Ein}). Considering relaxation of the charge density one can
ignore external field ${\bf E}$. The relaxation is described by
the continuity equation
\begin{equation}\label{c}
\frac{\partial(e n)}{\partial t}=-\nabla\cdot {\bf j}
\end{equation}
or
\begin{equation}
\label{3re}
 \frac{\partial(e n)}{\partial t}=\sigma\left(\frac{1}{e^2} \frac{d \mu}{d n}\nabla^2(e n^\prime)-\frac{4 \pi e
 n^\prime}{\kappa}\right).
\end{equation}
The ratio $R$ of the first (diffusion) term in the right hand side
to the second (field) term is $R=(q_3^2 L^2)^{-1}$, where
$L^{-2}=\nabla^2n^\prime/n^\prime$ is the characteristic  size of
the extra charge and $q_3^2$ is given by Eq. (\ref{3r}). If
electron gas is non-ideal, $q_3\sim 1/\bar{r}$,
 where $ \bar{r}$ is the average distance between electrons. However,
the very concept of diffusion equation is valid at $L\gg \bar{r}$.
This means that for
 the non-ideal gas $|R|\ll 1$ and the diffusion term in Eq. (\ref{3re})
 should be ignored. Then the equation has a simple solution
\begin{equation}\label{mrel}
n^\prime({\bf r},t)=n^\prime({\bf r},0)\exp -(t/\tau_M),
\end{equation}
where $\tau_M=\kappa/(4\pi\sigma)$ is well-known Maxwell's time.
 Coefficient $D$ does not enter
  in this case in the entropy production and it does not have a physical sense.
  Thus in 3-dimensional non-ideal electron gas negative $d\mu/dn$ does not
  create any contradiction with the Einstein relation.

In the 3d gas of high density $\mu \sim n^{2/3}$ and $R\sim
(\bar{r}/L)^2/r_s$ with $r_s<1$. In this case $R$ might be large
and diffusion is possible. However $d\mu/dn>0$, and $D>0$.

Now we consider the relaxation of the charge density in
2-dimensional case. Instead of Eq.(\ref{3re}) one gets
\begin{eqnarray}
\nonumber
 \frac{\partial(e n_2)}{\partial t}=\sigma_2 (\frac{1}{e^2} \frac{d \mu}{d n_2}\nabla^2(e n^\prime_2)
 \nonumber
 \\
 -\frac{e}{\bar{\kappa}}\nabla^2\int\frac{n^\prime_2({\bf r^\prime}) d^2 r^\prime}{|{\bf r-r^\prime}|}).
 \label{2re}
\end{eqnarray}
Here $n_2, \sigma_2$ and $\nabla$ are 2-dimensional density,
conductivity, and 2-dimensional gradient respectively. To consider
the ratio $R_2$ of the first (diffusion) term to the second
(field) term it is convenient to make the Fourier transformation.
Then one gets
\begin{equation}\label{2f}
\frac{ \partial( n_q)}{\partial t}=-\sigma_2(\frac{1}{e^2} \frac{d
\mu}{d n_2}q^2 n_q +\frac{2\pi q}{\bar{\kappa}}n_q),
\end{equation}
where $n_q$ is the Fourier transformation of $n_2^\prime$.

Now we find that the ratio of the first ( diffusion) term in the
right hand side of Eq. (\ref{2f}) to the second (field) term
$R_2=q/q_2$, where $q_2$ is given by Eq. (\ref{2r}). Similar to
the 3d case in the non-ideal gas $|q_2|\sim 1/\bar{r}$ and
diffusion should be ignored. Then we get the Dyakonov-Furman
equation\cite{DF}
\begin{equation}\label{DF}
\frac{\partial( n_q)}{\partial t}=-vqn_q,
\end{equation}
where velocity $v=2\pi\sigma_2/\bar{\kappa}$. The physical meaning
of this equation is that extra density of electrons localized
initially at some spot propagates in all directions with velocity
$v$ conserving the total amount of extra electrons. Of course,
this way of relaxation is more efficient than diffusion (random
walk), because $r\sim vt$ while $r\sim \sqrt{Dt}$   in the case of
diffusion. Thus, diffusion coefficient $D$ is irrelevant and
negative $d\mu/dn$ does not create any contradiction with the
Einstein relation In a high density electron gas $R_2=q
\bar{r}/r_s$ and diffusion mechanism is possible.  In this case
$d\mu/dn>0$ and $D>0$.

One can consider this problem from a different point of view. In
both 3d and 2d  cases the negative diffusion coefficient $D$
appears in the term with the highest derivative that leads to the absolute
instability even if $D$ is small\cite{rashba}. Consider, for
example Eq. (\ref{3re}) for 3d case. After the Fourier
transformation the solution for the charge density $\rho=e n^\prime$ can be
written in a form
\begin{equation}\label{f}
\rho_q=\rho_q^0\exp\left(-\frac{4\pi\sigma t}{\kappa}-D q^2 t\right),
\end{equation}
where $D$ is given by the Einstein relation Eq. (\ref{Ein}). One
can see that at $D<0$ solution increases with time exponentially
for harmonics with $q \bar{r}\geq 1$.

The physical explanation is as follows.  The
Eqs.(\ref{3re},\ref{2re})
 contain average distance between electrons $\bar{r}$.
So they contain information that the charged liquid has a discreet
electronic structure. This information comes from the negative density
of states which originates from the interaction of the separate electrons.
 That is why macroscopic equations become unstable at small spacial
 harmonics. The message is that $n({\bf r})$ is rather a set of
 $\delta$-functions than a continuous function.
The instability is absent if $D$ is positive.

The instability of small spatial harmonics at  small negative $D$
does not affect larger harmonics because Eqs.(\ref{3re},\ref{2re})
are linear. Due to the linearity different harmonics are
independent and transformation of energy from  small spacial
harmonics to large harmonics  is forbidden (cp. phenomenon of
turbulence
 in non-linear hydrodynamics where the transformation of energy is not forbidden,
 but the instability is initiated by large harmonics).

 Therefore, I think that  at small $D$ approximation $D=0$ that
gives Eqs.(\ref{mrel},\ref{DF}) is correct.

One should note that the problem of
the non-physical roots of electric permittivity discussed in the
previous section is of the same  nature.

Before we discussed the unipolar diffusion. Consider the simplest
case of the ambipolar diffusion
 assuming that  at $t=0$ the densities of electrons and holes are equal in
 some finite region of space and are zero otherwise. Moreover we assume that the local macroscopic
 charge density $\rho({\bf r},t)=0$ and a recombination of carriers is very slow.  In this case
Eq. (\ref{ec2}) describes the electron-hole system in
quasi-equilibrium. At large $r_s$ one gets $E, \mu, d\mu/dn <0$
but
 the last term in Eq. (\ref{ec2}) is absent. So the smearing of the density of particles is
 described by the equation of diffusion  at all $r_s$, but  at small density ($r_s\geq 1$)
  coefficient $D<0$. Then the absolute instability  takes place for all
  harmonics that means a collapse of the system. Thus the
  electron-hole "Wigner liquid" and  crystal are unstable.

   This result is
  very transparent. It happens because negative $\mu$ just means that the energy
  of the system decreases with increasing density. In bipolar case neutrality is
   provided by the particles and we do not consider any background. Thus the instability is a result of the negative compressibility in a neutral system. At large enough $r_s$ these
   particles are classical, and the absence of the mechanical equilibrium
  follows also from the Earnshaw theorem. In reality quantum mechanics
  becomes more
  important with increasing density. As a result the excitons  are formed. These  neutral
 particles have a positive diffusion coefficient $D_a$ and their
 density  smears with time through all available space. This
 process is described by a regular diffusion equation.
 In the case of optical excitation the carriers  may appear in
 the form of the excitons from the very beginning

 For the coefficient of the  ambipolar diffusion $D_a$ a textbook
 equation\cite{see}
\begin{equation}\label{ad}
D_a=\frac{2D_e D_h}{D_e+D_h}
\end{equation}
is often used, where $D_{e,h}$ are diffusion coefficients of
electron and holes in unipolar case. As follows from the previous
discussion, one should be careful with this equation because for
the non-ideal electron (or hole) gas these unipolar coefficients
might be negative and meaningless. It happens because in unipolar
case there is a deviation from neutrality that creates electric
field, while in bipolar case the system is neutral. In this case
Eq. (\ref{ad}) does not work and one should calculate $D_a$ in a
different way as a diffusion of the exciton.

In the recent paper by Zhao\cite{zhai} the experimental results
for the ambipolar diffusion in silicon-on-insulator system are
compared with Eq. (\ref{ad}). At high temperatures a good
agreement is found while at low temperatures the observed values
of $D_a$ are 6-7 times less. The previously reported
values\cite{pr} show similar temperature dependence.

The author's explanation is that coefficients $D_{e,h}$ are taken
for the bulk silicon using Einstein relation and they might be
larger than in the film at low temperatures. However, the reason
discussed above cannot be excluded.

\section{Graphene as a possible example of a non-ideal electron system}
 It is interesting to discuss the single layer graphene
as an example of the system with non-ideal electron gas.
 Graphene is a gapless
material with the linear spectrum of electrons and holes near the
Dirac point. Due to some reasons, that are not quite clear now,
the velocity $v$ of electrons and holes in equation $\epsilon=\pm
pv$ is of the order of $e^2/\hbar$. It follows  that at any Fermi
energy inside this linear spectrum electron gas in graphene is
non-ideal in a sense mentioned above: the absolute value of the
chemical potential  is of the order of interaction energy $e^2
n^{1/2}$. It means that unipolar density relaxation in this system
should be described by the Dyakonov-Furman equation rather than by
diffusion equation.

However,without magnetic field the electron gas in graphene is
marginally non-ideal. It cannot be classical, like an electron gas
of a low density with quadratic spectrum. The marginal situation
makes theoretical calculations very difficult. Nevertheless, it is
accepted that the Wigner crystal in single layer graphene is
absent without magnetic field\cite{balat,cote}. The sign of
$d\mu/dn$ is also an interesting question but very difficult for
theoretical study.   Recently tunneling microscopy experiment has
been done by Martin {\it et al.}\cite{mart}. They claim that their
measurement give the thermodynamic density of states and that it
is positive. The last statement  might be a result of disorder.
\section{Conclusion}
Finally I argue that the negative sign of diffusion coefficient
that follows from the Einstein relation at negative density of
states does not lead to any contradiction because diffusion
coefficient is irrelevant  for the unipolar transport under this
condition. The sign of the diffusion coefficient in this case
should not be definitely positive because the diffusion is not the
main source of the entropy production. In bipolar situation
negative diffusion means the collapse of the system and formation
of neutral excitons.

I am grateful to Boris Shklovskii and  Yoseph Imry  for important
discussion. I am especially indebted to  David Khmelnitskii and
Emmanuel Rashba for multiple discussions and criticism.

\bibliography{Einst}

\begin{thebibliography}{27}
\expandafter\ifx\csname natexlab\endcsname\relax\def\natexlab#1{#1}\fi
\expandafter\ifx\csname bibnamefont\endcsname\relax
  \def\bibnamefont#1{#1}\fi
\expandafter\ifx\csname bibfnamefont\endcsname\relax
  \def\bibfnamefont#1{#1}\fi
\expandafter\ifx\csname citenamefont\endcsname\relax
  \def\citenamefont#1{#1}\fi
\expandafter\ifx\csname url\endcsname\relax
  \def\url#1{\texttt{#1}}\fi
\expandafter\ifx\csname urlprefix\endcsname\relax\def\urlprefix{URL }\fi
\providecommand{\bibinfo}[2]{#2}
\providecommand{\eprint}[2][]{\url{#2}}

\bibitem[{\citenamefont{Einstein}(1905)}]{ALB}
\bibinfo{author}{\bibfnamefont{A.}~\bibnamefont{Einstein}},
  \bibinfo{journal}{Annalen der Physik} \textbf{\bibinfo{volume}{17}},
  \bibinfo{pages}{549} (\bibinfo{year}{1905}).

\bibitem[{\citenamefont{von Smoluchowsky}(1906)}]{SM}
\bibinfo{author}{\bibfnamefont{M.}~\bibnamefont{von Smoluchowsky}},
  \bibinfo{journal}{Annalen der Physik} \textbf{\bibinfo{volume}{21}},
  \bibinfo{pages}{756} (\bibinfo{year}{1906}).

\bibitem[{\citenamefont{Landau and Lifshitz}(1984)}]{LL}
\bibinfo{author}{\bibfnamefont{L.~D.} \bibnamefont{Landau}} \bibnamefont{and}
  \bibinfo{author}{\bibfnamefont{E.}~\bibnamefont{Lifshitz}},
  \emph{\bibinfo{title}{Electrodynamics of Continuous Media}}
  (\bibinfo{publisher}{Butterworth-Heinenann}, \bibinfo{year}{1984}),
  \bibinfo{note}{chapter III}.

\bibitem[{\citenamefont{Altshuler et~al.}(1980)\citenamefont{Altshuler, Aronov,
  and Lee}}]{alt}
\bibinfo{author}{\bibfnamefont{B.~L.} \bibnamefont{Altshuler}},
  \bibinfo{author}{\bibfnamefont{A.~G.} \bibnamefont{Aronov}},
  \bibnamefont{and} \bibinfo{author}{\bibfnamefont{P.~A.} \bibnamefont{Lee}},
  \bibinfo{journal}{Phys. Rev. Lett} \textbf{\bibinfo{volume}{44}},
  \bibinfo{pages}{1288} (\bibinfo{year}{1980}).

\bibitem[{\citenamefont{Zyuzin}(1981)}]{zu}
\bibinfo{author}{\bibfnamefont{A.~Y.} \bibnamefont{Zyuzin}},
  \bibinfo{journal}{JETP Lett.} \textbf{\bibinfo{volume}{33}},
  \bibinfo{pages}{360} (\bibinfo{year}{1981}).

\bibitem[{\citenamefont{Altshuler and Aronov}(1985)}]{alt1}
\bibinfo{author}{\bibfnamefont{B.~L.} \bibnamefont{Altshuler}}
  \bibnamefont{and} \bibinfo{author}{\bibfnamefont{A.~G.}
  \bibnamefont{Aronov}}, \emph{\bibinfo{title}{Electron-Electron Interaction in
  Disordered Systems (ed. by A. L. Efros and M. Pollak)}}
  (\bibinfo{publisher}{North-Holland, Amsterdam}, \bibinfo{year}{1985}),
  \bibinfo{note}{p.\ 37}.

\bibitem[{\citenamefont{Shklovskii and Efros}(1987)}]{SE}
\bibinfo{author}{\bibfnamefont{B.~I.} \bibnamefont{Shklovskii}}
  \bibnamefont{and} \bibinfo{author}{\bibfnamefont{A.~L.} \bibnamefont{Efros}},
  \bibinfo{journal}{JETP Lett.} \textbf{\bibinfo{volume}{44}},
  \bibinfo{pages}{669} (\bibinfo{year}{1987}).

\bibitem[{\citenamefont{Bello et~al.}(1981)\citenamefont{Bello, Levin,
  Shklovskii, and Efros}}]{BE}
\bibinfo{author}{\bibfnamefont{M.~S.} \bibnamefont{Bello}},
  \bibinfo{author}{\bibfnamefont{E.~I.} \bibnamefont{Levin}},
  \bibinfo{author}{\bibfnamefont{B.~I.} \bibnamefont{Shklovskii}},
  \bibnamefont{and} \bibinfo{author}{\bibfnamefont{A.~L.} \bibnamefont{Efros}},
  \bibinfo{journal}{Sov. Phys JETP} \textbf{\bibinfo{volume}{53}},
  \bibinfo{pages}{822} (\bibinfo{year}{1981}).

\bibitem[{\citenamefont{Kravchenko et~al.}(1989)\citenamefont{Kravchenko,
  Pudalov, and Semenchinsky}}]{KRL}
\bibinfo{author}{\bibfnamefont{S.~V.} \bibnamefont{Kravchenko}},
  \bibinfo{author}{\bibfnamefont{V.~M.} \bibnamefont{Pudalov}},
  \bibnamefont{and} \bibinfo{author}{\bibfnamefont{S.~G.}
  \bibnamefont{Semenchinsky}}, \bibinfo{journal}{Phys. Lett. A}
  \textbf{\bibinfo{volume}{141}}, \bibinfo{pages}{71} (\bibinfo{year}{1989}).

\bibitem[{\citenamefont{Kravchenko et~al.}(1990)\citenamefont{Kravchenko,
  Rinberg, Semenchinsky, and Pudalov}}]{KR}
\bibinfo{author}{\bibfnamefont{S.~V.} \bibnamefont{Kravchenko}},
  \bibinfo{author}{\bibfnamefont{D.~A.} \bibnamefont{Rinberg}},
  \bibinfo{author}{\bibfnamefont{S.~G.} \bibnamefont{Semenchinsky}},
  \bibnamefont{and} \bibinfo{author}{\bibfnamefont{V.~M.}
  \bibnamefont{Pudalov}}, \bibinfo{journal}{Phys. Rev. B}
  \textbf{\bibinfo{volume}{42}}, \bibinfo{pages}{3741} (\bibinfo{year}{1990}).

\bibitem[{\citenamefont{Eisenstein et~al.}(1992)\citenamefont{Eisenstein,
  Pfeiffer, and West}}]{EI}
\bibinfo{author}{\bibfnamefont{J.~P.} \bibnamefont{Eisenstein}},
  \bibinfo{author}{\bibfnamefont{L.~N.} \bibnamefont{Pfeiffer}},
  \bibnamefont{and} \bibinfo{author}{\bibfnamefont{K.}~\bibnamefont{West}},
  \bibinfo{journal}{Phys. Rev. Lett} \textbf{\bibinfo{volume}{68}},
  \bibinfo{pages}{674} (\bibinfo{year}{1992}).

\bibitem[{\citenamefont{Eisenstein et~al.}(1994)\citenamefont{Eisenstein,
  Pfeiffer, and West}}]{EI1}
\bibinfo{author}{\bibfnamefont{J.~P.} \bibnamefont{Eisenstein}},
  \bibinfo{author}{\bibfnamefont{L.~N.} \bibnamefont{Pfeiffer}},
  \bibnamefont{and} \bibinfo{author}{\bibfnamefont{K.}~\bibnamefont{West}},
  \bibinfo{journal}{Phys. Rev. B} \textbf{\bibinfo{volume}{50}},
  \bibinfo{pages}{1760} (\bibinfo{year}{1994}).

\bibitem[{\citenamefont{Ando et~al.}(1982)\citenamefont{Ando, B.Fowler, and
  Stern}}]{ando}
\bibinfo{author}{\bibfnamefont{T.}~\bibnamefont{Ando}},
  \bibinfo{author}{\bibfnamefont{A.}~\bibnamefont{B.Fowler}}, \bibnamefont{and}
  \bibinfo{author}{\bibfnamefont{F.}~\bibnamefont{Stern}},
  \bibinfo{journal}{Rev.\ Mod.\ Phys.} \textbf{\bibinfo{volume}{54}},
  \bibinfo{pages}{437} (\bibinfo{year}{1982}).

\bibitem[{\citenamefont{Efros}(1992{\natexlab{a}})}]{ef1}
\bibinfo{author}{\bibfnamefont{A.~L.} \bibnamefont{Efros}},
  \bibinfo{journal}{Phys. Rev B} \textbf{\bibinfo{volume}{45}},
  \bibinfo{pages}{11354} (\bibinfo{year}{1992}{\natexlab{a}}).

\bibitem[{\citenamefont{Luryi}(1988)}]{SL}
\bibinfo{author}{\bibfnamefont{S.}~\bibnamefont{Luryi}},
  \bibinfo{journal}{Appl. Phys. Lett.} \textbf{\bibinfo{volume}{52}},
  \bibinfo{pages}{501} (\bibinfo{year}{1988}).

\bibitem[{\citenamefont{Efros et~al.}(1992)\citenamefont{Efros, Pikus, and
  Burnett}}]{BU}
\bibinfo{author}{\bibfnamefont{A.~L.} \bibnamefont{Efros}},
  \bibinfo{author}{\bibfnamefont{F.~G.} \bibnamefont{Pikus}}, \bibnamefont{and}
  \bibinfo{author}{\bibfnamefont{V.~G.} \bibnamefont{Burnett}},
  \bibinfo{journal}{Solid State Comm.} \textbf{\bibinfo{volume}{84}},
  \bibinfo{pages}{91} (\bibinfo{year}{1992}).

\bibitem[{\citenamefont{Pikus and Efros}(1993)}]{PE}
\bibinfo{author}{\bibfnamefont{F.~G.} \bibnamefont{Pikus}} \bibnamefont{and}
  \bibinfo{author}{\bibfnamefont{A.~L.} \bibnamefont{Efros}},
  \bibinfo{journal}{Phys. Rev. B} \textbf{\bibinfo{volume}{47}},
  \bibinfo{pages}{16395} (\bibinfo{year}{1993}).

\bibitem[{\citenamefont{Efros}(1992{\natexlab{b}})}]{AL}
\bibinfo{author}{\bibfnamefont{A.~L.} \bibnamefont{Efros}},
  \bibinfo{journal}{Phys. Rev. B} \textbf{\bibinfo{volume}{45}},
  \bibinfo{pages}{11354} (\bibinfo{year}{1992}{\natexlab{b}}).

\bibitem[{\citenamefont{Smith et~al.}(1986)\citenamefont{Smith, Wang, and
  Stiles}}]{S}
\bibinfo{author}{\bibfnamefont{T.~P.} \bibnamefont{Smith}},
  \bibinfo{author}{\bibfnamefont{W.~I.} \bibnamefont{Wang}}, \bibnamefont{and}
  \bibinfo{author}{\bibfnamefont{P.~J.} \bibnamefont{Stiles}},
  \bibinfo{journal}{Phys. Rev. B} \textbf{\bibinfo{volume}{34}},
  \bibinfo{pages}{2995} (\bibinfo{year}{1986}).

\bibitem[{\citenamefont{Dyakonov and Furman}(1987)}]{DF}
\bibinfo{author}{\bibfnamefont{M.~I.} \bibnamefont{Dyakonov}} \bibnamefont{and}
  \bibinfo{author}{\bibfnamefont{A.~S.} \bibnamefont{Furman}},
  \bibinfo{journal}{Sov. Phys. JETP} \textbf{\bibinfo{volume}{65}},
  \bibinfo{pages}{574} (\bibinfo{year}{1987}).

\bibitem[{ras()}]{rashba}
\bibinfo{note}{I am grateful to E. I. Rashba for this comment}.

\bibitem[{\citenamefont{Seeger}(1999)}]{see}
\bibinfo{author}{\bibfnamefont{K.}~\bibnamefont{Seeger}},
  \emph{\bibinfo{title}{Semiconductor Physics. An Introduction}}
  (\bibinfo{publisher}{Springer}, \bibinfo{year}{1999}), \bibinfo{note}{p.\
  124}.

\bibitem[{\citenamefont{Zhao}(2008)}]{zhai}
\bibinfo{author}{\bibfnamefont{H.}~\bibnamefont{Zhao}}, \bibinfo{journal}{Appl.
  Phys. Lett.} \textbf{\bibinfo{volume}{92}}, \bibinfo{pages}{112104}
  (\bibinfo{year}{2008}).

\bibitem[{\citenamefont{Rosling et~al.}(1994)\citenamefont{Rosling, Bleichner,
  Jonsson, and Nordlander}}]{pr}
\bibinfo{author}{\bibfnamefont{M.}~\bibnamefont{Rosling}},
  \bibinfo{author}{\bibfnamefont{H.}~\bibnamefont{Bleichner}},
  \bibinfo{author}{\bibfnamefont{P.}~\bibnamefont{Jonsson}}, \bibnamefont{and}
  \bibinfo{author}{\bibfnamefont{E.}~\bibnamefont{Nordlander}},
  \bibinfo{journal}{Appl. Phys. Lett.} \textbf{\bibinfo{volume}{76}},
  \bibinfo{pages}{2855} (\bibinfo{year}{1994}).

\bibitem[{\citenamefont{Dahal et~al.}()\citenamefont{Dahal, Wehling, Bedell,
  Zhu, and V.Balatsky}}]{balat}
\bibinfo{author}{\bibfnamefont{H.~P.} \bibnamefont{Dahal}},
  \bibinfo{author}{\bibfnamefont{T.~O.} \bibnamefont{Wehling}},
  \bibinfo{author}{\bibfnamefont{K.~S.} \bibnamefont{Bedell}},
  \bibinfo{author}{\bibfnamefont{J.-X.} \bibnamefont{Zhu}}, \bibnamefont{and}
  \bibinfo{author}{\bibfnamefont{A.}~\bibnamefont{V.Balatsky}},
  \bibinfo{note}{arXiv:cond-mat/0706.1689}.

\bibitem[{\citenamefont{Cote et~al.}()\citenamefont{Cote, Jobidon, and
  Fertig}}]{cote}
\bibinfo{author}{\bibfnamefont{R.}~\bibnamefont{Cote}},
  \bibinfo{author}{\bibfnamefont{J.-F.} \bibnamefont{Jobidon}},
  \bibnamefont{and} \bibinfo{author}{\bibfnamefont{H.~A.}
  \bibnamefont{Fertig}}, \bibinfo{note}{arXiv:cond-mat/0806.0573}.

\bibitem[{\citenamefont{Martin et~al.}(2008)\citenamefont{Martin, Akrman,
  Ulbricht, Lohmann, Smet, von Klitzing, and Yacoby}}]{mart}
\bibinfo{author}{\bibfnamefont{J.}~\bibnamefont{Martin}},
  \bibinfo{author}{\bibfnamefont{N.}~\bibnamefont{Akrman}},
  \bibinfo{author}{\bibfnamefont{G.}~\bibnamefont{Ulbricht}},
  \bibinfo{author}{\bibfnamefont{T.}~\bibnamefont{Lohmann}},
  \bibinfo{author}{\bibfnamefont{J.~H.} \bibnamefont{Smet}},
  \bibinfo{author}{\bibfnamefont{K.}~\bibnamefont{von Klitzing}},
  \bibnamefont{and} \bibinfo{author}{\bibfnamefont{A.}~\bibnamefont{Yacoby}},
  \bibinfo{journal}{Nature Physics} \textbf{\bibinfo{volume}{4}},
  \bibinfo{pages}{144} (\bibinfo{year}{2008}).

\end{thebibliography}
\end{document}